\documentclass[journal]{IEEEtran}
\ifCLASSINFOpdf
\usepackage[pdftex]{graphicx}
\usepackage{amsfonts}
\graphicspath{{../images/}}
\DeclareGraphicsExtensions{.pdf,.jpeg,.png}
\else
\fi

\usepackage[table]{xcolor}
\usepackage{algorithm,algorithmic}
\usepackage{amsmath}

\begin{document}

\title{High-Resolution Programmable Scattering\\ for Wireless Coverage Enhancement:\\ An Indoor Field Trial Campaign}

%\author{{James Rains, Jalil ur Rehman Kazim, Anvar Tukmanov, Lei Zhang,\\ Qammer H. Abbasi, Muhammad Imran, \& Tie Jun Cui}% <-this % stops a space
%\thanks{James Rains' PhD is funded by EPSRC ICASE studentship (EP/V519686/1) with British Telecom}}

 \author{James~Rains,~\IEEEmembership{Graduate Student Member,~IEEE,}
      Jalil~ur~Rehman~Kazim,~\IEEEmembership{Graduate Student Member,~IEEE,}\\
      Anvar~Tukmanov,~\IEEEmembership{Senior Member,~IEEE,}
      Tie~Jun~Cui,~\IEEEmembership{Fellow,~IEEE,}
      Lei~Zhang,~\IEEEmembership{Senior Member,~IEEE,}\\
      Qammer H. Abbasi,~\IEEEmembership{Senior Member,~IEEE,}
       and~Muhammad Ali Imran,~\IEEEmembership{Senior Member,~IEEE}}% <-this % stops a space
\maketitle

\begin{abstract}
  This paper presents a multi-bit reconfigurable intelligent surface (RIS) with a high phase resolution, capable of beam-steering in the azimuthal plane at sub-6 Gigahertz (GHz). \color{black} Field trials in realistic indoor deployments have been carried out, with coverage enhancement performance ascertained for three common wireless communication scenarios. Namely, serving users in an open lobby with mixed line of sight and non-line of sight conditions, communication via a junction between long corridors, and a multi-floor scenario with propagation via windows. This work explores the potential for reconfigurable intelligent surface (RIS) deployment to mitigate non-line of sight effects in indoor wireless communications. In a single transmitter, single receiver non-line of sight link, received power improvement of as much as 40 dB is shown to be achievable by suitable placement of a RIS, with an instantaneous bandwidth of at least 100 MHz possible over a 3 to 4.5 GHz range. In addition, the effects of phase resolution on the optimal power reception for the multi-bit RIS have been experimentally verified, with a 2.65 dB improvement compared to a 1-bit case. 
\end{abstract}

% Note that keywords are not normally used for peer review papers.
\begin{IEEEkeywords}
Reconfigurable Intelligent Surfaces, Intelligent Reflecting Surfaces, Programmable Metasurfaces, Wireless Coverage, Smart Radio Environments
\end{IEEEkeywords}

% \fi
\IEEEpeerreviewmaketitle

\section{Introduction}

%\linenumbers

\IEEEPARstart{R}{econfigurable} wireless propagation environments may become a key enabler of ubiquitous reliable communications in future mobile networks. Recent advancements in reconfigurable metasurfaces have shown promising electromagnetic (EM) wave transformation capability without the need for complex and high power-consuming electronics \cite{Cui2017}\cite{Zhang2019a}. The required nature of these wave transformations is subject to a great deal of research, with the distinction between what is possible and what is useful still open research problem. Reconfigurable intelligent surfaces (RISs) have recently achieved significant attention from wireless communications researchers and industry due to their programmable beam focusing and anomalous reflection capability \cite{Gros2021}, with the European Telecommunications Standards Institute (ETSI) recently launching an industry specification group (ISG) for RISs to begin standardisation efforts. 

\subsection{Reconfigurable Intelligent Surface Architecture}

RISs are essentially reflecting-type reconfigurable metasurfaces with a means of being programmed via a control link. A RIS can selectively excite spatial harmonics from incident EM waves by varying the periodicity of the phase-gradient presented by its constituent unit cell elements. These two-dimensional structures could easily be integrated into our built environment, to extend the coverage of mobile base stations to blind spots as well as selectively improving achievable data rates at cell edges \cite{Renzo2020}\cite{Bjoernson2021}.

% Intro to RIS architecture
Reflecting-type reconfigurable metasurfaces typically consist of hundreds of unit cell elements made up of microstrip patches printed on a dielectric substrate, with a thickness below $\lambda/10$ \cite{Cui2014}, backed by a ground plane. The periodicity of the unit cells is usually less than $\lambda/2$ in order to ensure excitation of parasitic spatial harmonics is kept to a minimum \cite{Munk2000}. In the microwave region, the local reflection coefficient of a RIS element can be varied by placing a tunable load between sets of microstrip patches, such as a varactor diode or a positive-intrinsic-negative (PIN) diode switch \cite{Zhang2019}\cite{Pei2021}. PIN diodes are desirable for their low voltage requirements, enabling interfacing with off-the-shelf shift registers. At sub-6 GHz, PIN diodes are utilised as switches, with forward-biased states acting as a series resistance of a few ohms and a reverse-biased state with a series capacitance typically below 1 pF. A single PIN diode can be employed to realise a single pair of local reflection coefficients, which can be represented by a single, 1-bit, binary state. A varactor diode interfaced with voltage level shifter circuitry can be utilised to the same effect \cite{Pei2021}. A greater number of unit cell reflection states can be realised by employing more PIN diodes per unit cell or interfacing varactor diodes with digital to analog converters (DACs) \cite{Zhu2013}. The former can be realised with more shift registers and results in higher current consumption per unit cell, whereas the later requires more expensive biasing circuitry. The application-dependent trade offs of unit cell resolution versus control circuitry complexity and power consumption is an open research area. 

% Where RISs have been tested thus far - give a summary of the field trials

\subsection{Reconfigurable Intelligent Surface Field Trials}
In the last few years, details on several RIS experimental testbeds have been published, with experiments performed in a diverse set of scenarios at microwave and millimetre-wave (mmWave) bands \cite{Gros2021}\cite{Pei2021}\cite{Trichopoulos2021}\cite{Popov2021}. 

% Pei 
Pei et al. \cite{Pei2021} performed indoor and long-range outdoor field trials with a varactor diode-based 1-bit RIS consisting of 1100 unit cell elements in a 20 row by 55 column arrangement. Operating at 5.8 GHz, the lateral dimensions of the RIS were 79 cm ($15.2 \lambda$) in width and 31 cm ($6 \lambda$) in height. To reduce feed network complexity and power consumption, unit cells were grouped in columns consisting of 5 unit cells each. A greedy fast beamforming algorithm was introduced, iteratively increasing the received power at a single receiver (Rx) antenna from a single-antenna transmitter (Tx), facilitated by an Rx-RIS feedback loop. In an indoor non-line of sight (NLoS) scenario, a receive horn antenna and the RIS formed a reflectarray-type arrangement and the received power was iteratively increased from a transmitter placed behind a thick concrete wall. Compared to a copper plate of similar dimensions to the RIS, the received power improvement was as much as 26 dB. Experiments performed on a rooftop, with a receiver horn antenna placed within a few meters from the RIS and a transmitter located at 50 m and 500 m. Thus, with the Tx-RIS-Rx forming a virtual line of sight (VLoS) link, the authors demonstrated received power improvements of 27 dB and 14 dB, respectively with a power consumption of only 1W.  Although the rooftop trials in this work did not involve realistic blockages, a directional antenna on the receiver ensured a limited direct path between the transmitter and receiver. 

Trichopoulos et al. \cite{Trichopoulos2021} introduced a monoplanar reconfigurable metasurface operating at 5.8 GHz with 160 individually-addressable 1-bit PIN diode-based unit cells in a 10 row by 16 column arrangement. The lateral dimensions of the prototype were 41.4 cm ($8\lambda$) in width and 25.9 cm ($5\lambda$) in height. The VIAless approach of this design, as well as a wide unit cell periodicity of $\lambda/2$ ensures scalability to operation at mmWave, at the expense of additional parasitic spatial modes. The authors utilised a codebook-based beamforming algorithm. A realistic outdoor measurement campaign was performed in a line-of-sight scenario and a scenario with significant blockage between the transmitter and receiver antennas. The authors mapped coverage improvement for reception from an omnidirectional antenna placed in the blockage region 35 meters from the RIS. Compared to the case of no RIS, average received power improvement of 6 dB and a maximum improvement of 8 dB was demonstrated.

Operating at mmWave bands presents challenges of high susceptibility to blockages, limiting reliable communication to LoS scenarios. RISs may provide significant coverage enhancement at mmWave for their ease of scalability means that they can provide large effective aperture gains via VLoS paths which can follow the user equipment in the region of interest. Greenerwave have been commercially developing PIN diode-based dual-polarised 1-bit reconfigurable metasurfaces for use as passive access point extenders at mmWave frequencies, as is detailed in the work by Gros et al. \cite{Gros2021}. The authors introduced a 400-element RIS in a 20 by 20 arrangement with lateral dimensions of 10cm $\times$ 10 cm ($\approx 10\lambda \times 10 \lambda$) with a centre operating frequency of 28.5 GHz. In their experiment, a blockage is created with a right-angled barrier which could equally be the corner of a building. Introducing a VLoS link via the outside corner of the barrier resulted in a 25 dB improvement compared to a copper plate of similar dimensions, with a 3dB bandwidth of over 250 MHz. A 1600 element (40 $\times$ 40) version of this RIS design was also utilised by Popov et al. in \cite{Popov2021} as a passive range extender to serve a user in a room from a transmitter antenna in a corridor. The RIS was placed in the corridor on a wall opposite the room doorway, forming a VLoS link and realising a 30 dB received power improvement at 29.5 GHz over a 3 GHz bandwidth, with a power consumption below 6W.  

Most published RIS experimental testbeds have only considered a 1-bit individually-addressable unit cell design due to the low complexity and small configuration overhead compared to higher resolution designs. Oftentimes in indoor communication scenarios, user equipment is co-located in the same horizontal plane. In these cases, beamsteering capability in the elevation is underutilised. Similarly, when network equipment is situated at distances outside of the radiative near-field of the RIS, as well as in the same horizontal plane, the optimal RIS configuration often resembles columns of unit cells with similar reflection states. For the 1-bit RIS case, this usually results in the generation of spurious beams, therefore wasting transmitted power and the potential to cause interference with nearby user equipment. Beam pattern purity is greatly enhanced for 2-bit and 3-bit designs \cite{Wu2008}. By utilising the column-connected approach to RIS design, elevation beam steering capability has been exchanged for a high reflection state resolution, thereby enabling performance enhancement in indoor communication scenarios without significant additional control circuit complexity. 

 To keep this investigation relevant to current and soon to be deployed wireless communications technologies, we opted to design a RIS operating at sub-6 GHz. In this region, the n77 5G New Radio (NR) 3.3 – 4.2 GHz band is particularly important for European mobile operators \cite{Schumacher2019}. %Alongside this, the frequency region 3.7 – 4.2 GHz, which has historically been allocated for fixed receive-only satellite earth stations, is in the process of being allocated for shared and private use \cite{Hattab2018}.
In this work, we investigate wireless channel improvement performance over 3 – 4.5 GHz to encompass this bands whilst obtaining results valid for potential future expansion of the 5G mid-band spectrum. The 5G testbed at University of Glasgow is licensed for use at 3.75 GHz and, to ensure compatibility with future measurement campaigns, the unit cell reflection responses were optimised centred at that frequency point.

\color{black}
Our contributions to the literature are as follows. Firstly, we have introduced a multi-bit RIS design with 7 distinct phase shifts for operation at 3.75 GHz and verified its global reflection performance. Secondly, we document here results of a measurement campaign with the fabricated RIS in three common indoor communication scenarios. Namely, serving users in an open lobby with mixed line of sight and non-line of sight conditions, communication via a junction between long corridors, and a multi-floor scenario with propagation via windows. Thirdly, we have verified and compared, in the field, received power enhancement in a complex propagation environment via increased phase resolution.   

\section{Multi-Bit Reconfigurable Metasurface}

\subsection{Unit Cell Design}

\begin{figure}
\centerline{\includegraphics[scale=0.11]{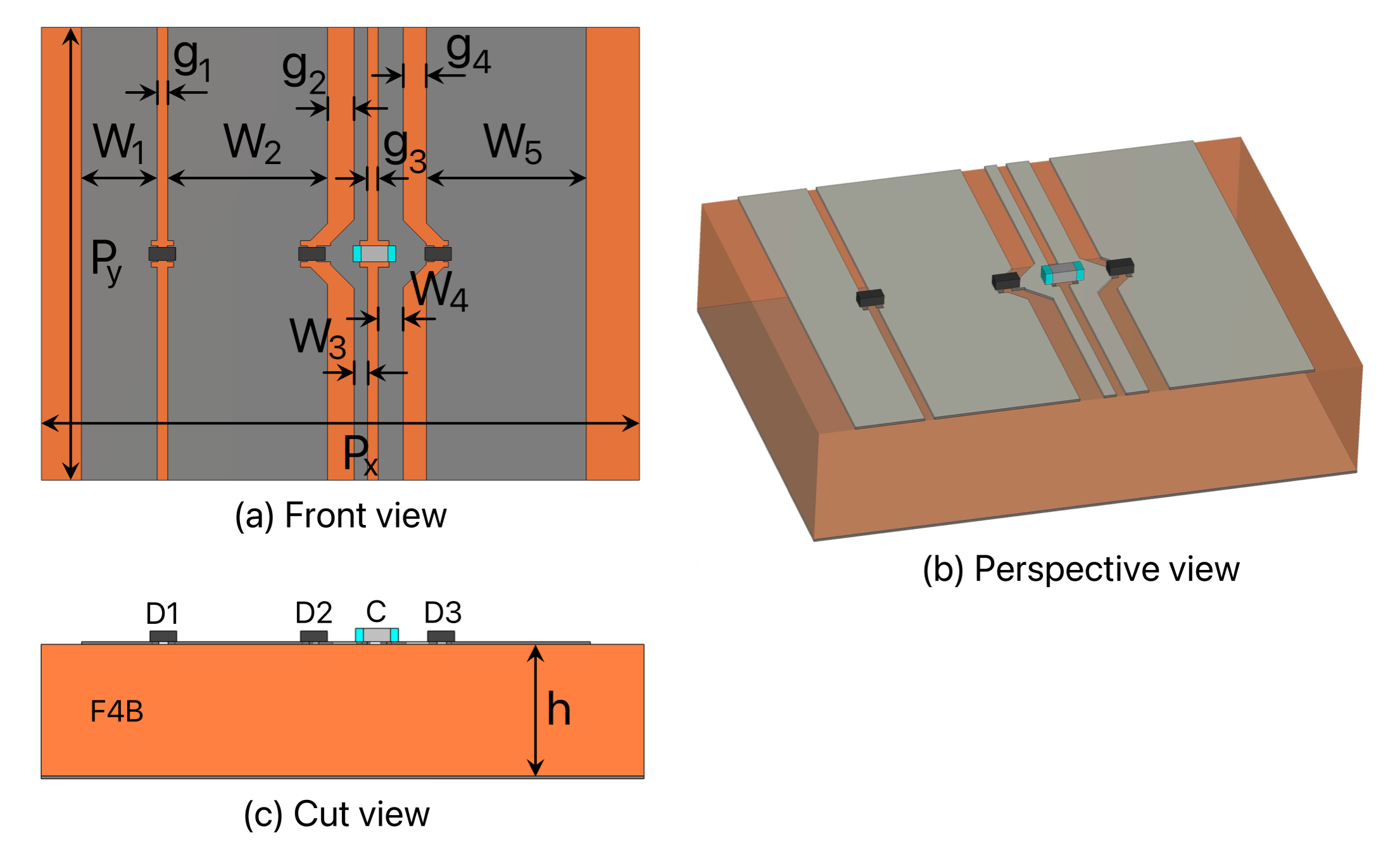}}
\caption{Multi-bit unit cell design used in this work. The unit cell consists of 5 rectangular patches on an F4B substrate, connected by 3 PIN diodes and a capacitor. }
\label{unitcell}
\end{figure}

The unit cell employed in this work is the multi-bit column-driven planar design depicted in Fig. \ref{unitcell}, which was recently introduced by the authors \cite{Rains21}. Each unit cell consists of 5 patches connected by 3 PIN diodes and a capacitor, mounted on a 5mm PTFE-based F4BM-2 substrate with relative permittivity $\epsilon_{r} = 2.65$ and loss tangent $tan\delta = 0.001$. Unit cells are biased in a column-wise fashion by applying DC voltages to 3 of the patches, with the remaining two patches connected to DC ground. The patch widths and spacing were optimised through a particle swarm optimisation algorithm in order to maximise the achievable phase resolution  whilst minimising the average reflection loss. For further elaboration on the geometry optimisation procedure, see \cite{Rains21}. \color{black} The components used in this unit cell design are Skyworks SMP1321-040LF PIN diodes and an AVX U-Series 3.6 pF 0402 capacitor. In order to increase the accuracy of simulations during design, the S-parameter data from the manufacturers of the PIN diodes and capacitors were incorporated in the simulations. The logic 0 state was simulated as that of 0 V PIN diode reverse bias voltage, whilst the logic 1 state refers to a forward biased condition with a current of 3 mA. 

\begin{table}[!t]
\renewcommand{\arraystretch}{1.3}
\caption{Dimensions for Multi-Bit Unit Cell Design}
\label{table_dims}
\centering
\begin{tabular}{|l|l|}
\hline
\hline
\multicolumn{1}{|c|}{Parameter} &
\multicolumn{1}{c|}{Dimensions (mm)} \\
\hline
Periodicity, $P_{x}$ $P_{y}$ & 22.5 15.0 \\
\hline
Patch width, $W_{1}$ to $W_{5}$ & 6.0 0.9 0.5 6.0 2.9  \\
\hline
Patch spacing, $g_{1}$ to $g_{4}$ & 0.9 0.4 1.0 0.4  \\
\hline
Substrate thickness, $h$ & 5.0  \\
\hline
\hline
\end{tabular}\end{table}

\begin{figure}
\centerline{\includegraphics[scale=0.6]{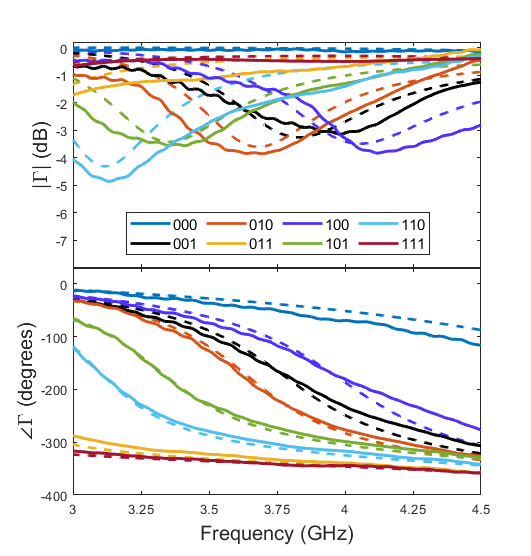}}
\caption{Simulated (dashed curves) and measured (solid curves) reflection response for 8 digital states of 3-bit reflecting metasurface.}
\label{phase_response}
\end{figure}

Compared to unit cell designs with VIAs, the planar structure of this design is desirable for its low fabrication complexity and wide bandwidth. This approach, however, could make maintaining signal integrity difficult due to the inflexibility of the shape of the bias feed traces (i.e., the columns of connected patches), particularly for applications that might require high-speed switching such as with recently introduced space-time modulation schemes \cite{Tang2020}. With a change in the type of PIN diodes employed and modifications to the pad spacing, this design is easily scalable for mmWave applications. 

\begin{figure}
\centerline{\includegraphics[scale=0.16]{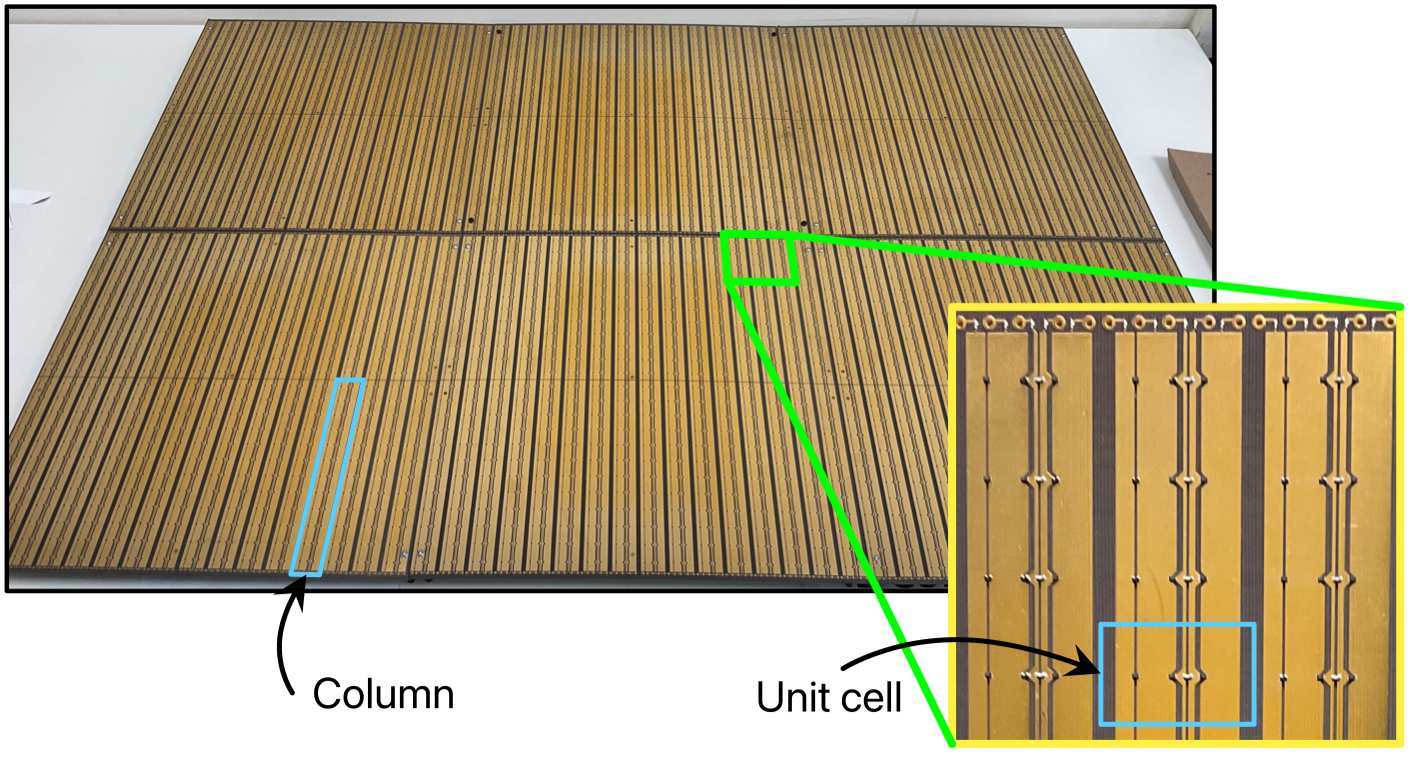}}
\caption{Fabricated reconfigurable intelligent surface. Inset shows column structure with choke inductors at the top. Column highlighted consists of a group of 12 unit cells whose patches are connected top and bottom.}
\label{fabricated}
\end{figure}

The fabricated RIS can be seen in Fig. \ref{fabricated}. Due to fabrication constraints on the lateral dimensions, the RIS was split into 6 tiles, each containing an arrangement of 16 $\times$ 24 unit cells. The lateral dimensions of this design are 1.08 m ($13.5\lambda$) in width and 0.72 m ($9\lambda$) in height. The total number of addressable columns on each tile is 32 (i.e., two rows of 16 columns), each controlled by 3 digital values for a total of 96 bits per tile. Accordingly, each tile contains 12 shift registers, which are interfaced by an external controller consisting of an FPGA and Raspberry Pi single-board computer. The control link is maintained over a 2.4 GHz WiFi connection between a PC and the Raspberry Pi.

The PIN diodes require a bias voltage of 0.83 V at the design current of 3 mA. On a per-unit cell basis, this voltage is easily accommodated by off-the-shelf shift registers. However, source drivers were necessary in order to supply enough current to drive 12 PIN diodes per digital output. These are connected in series with the shift register outputs. At full load (i.e., all outputs at digital high), each source driver consisting of 8 output pins is required to source 288 mA. Assuming on average half of the PIN diodes are on at any one moment, the PIN diodes alone consume 8.6 W. By no means does this meet the definition of ultra-low power required to justify the deployment of RISs, but this does not pose limitations on the investigation of the benefits of a higher phase resolution presented here. Varactor diodes should be employed if low power consumption is required, such as the testbed recently demonstrated by Pei et al. \cite{Pei2021}.

% the goal of this work was to experimentally verify potential performance enhancement of higher phase resolution digital metasurfaces

\begin{figure}[htbp]
\centerline{\includegraphics[scale=0.07]{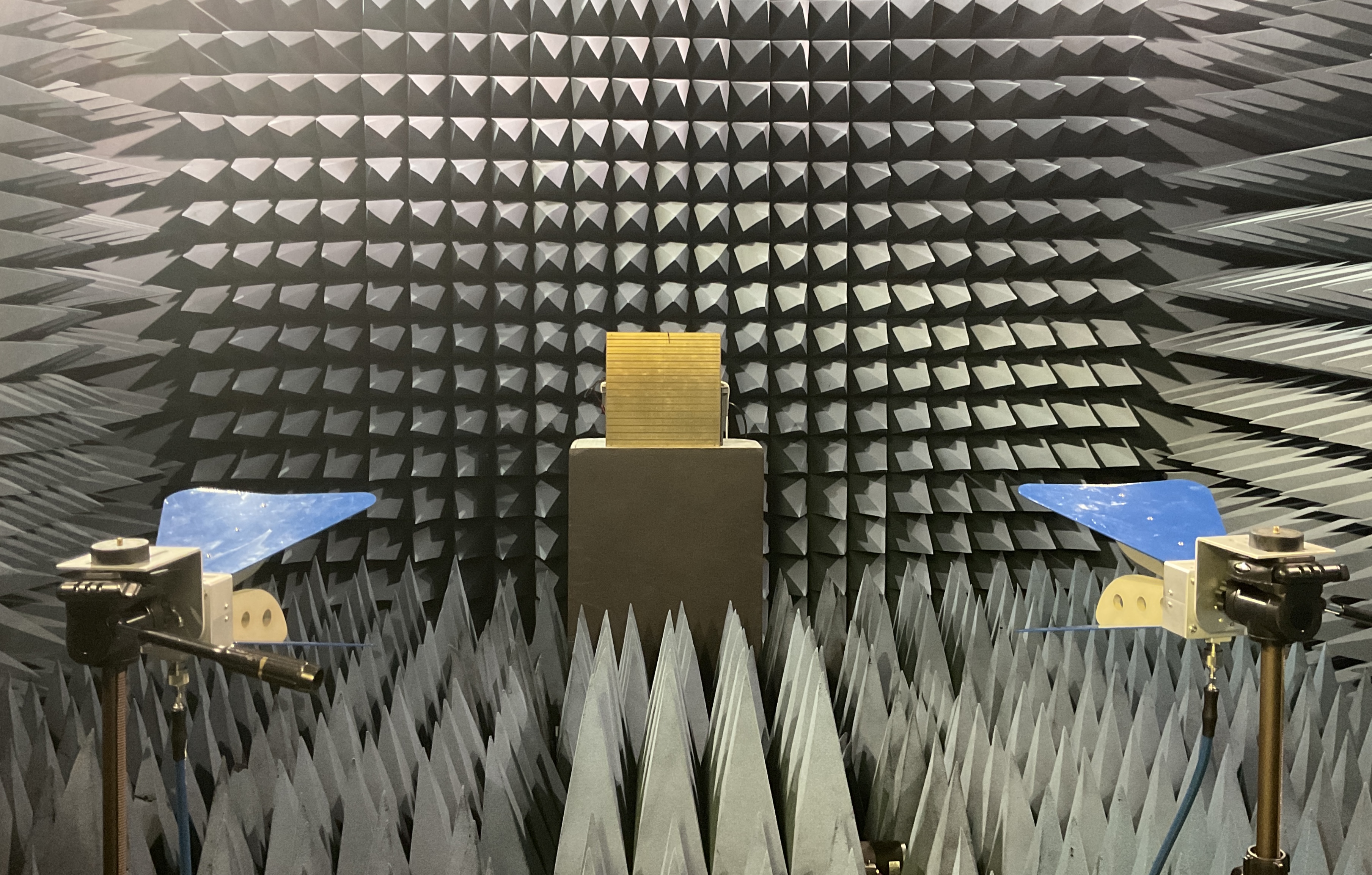}}
\caption{Measurement setup in an anechoic chamber for ascertaining global reflection characteristics. Two identical standard gain horn antennas in a horizontal polarisation configuration are placed in the horizontal plane from the center of a single RIS tile. The antennas are placed at 4 meters, 10 degrees from normal to the RIS and aligned to the RIS center with a laser pointer and spirit level. A Rohde and Schwarz ZVH8 VNA is connected to the horn antennas for the S21 measurements, with calibration performed by replacing the RIS with an aluminium plate of similar dimensions. }
\label{chamberRIS}
\end{figure}

\color{black}

\subsection{Performance Analysis}

A single RIS tile was placed in an anechoic chamber to ascertain its global reflection characteristics, as shown in Fig. \ref{chamberRIS}. The tile was oriented such that the columns of patches were horizontally arranged. Two identical standard gain horn antennas in a vertical polarisation configuration were placed in the horizontal plane from the center of the RIS tile. The antennas were placed at a distance of 4 meters, 10 degrees from normal to the RIS and aligned to the RIS center with a laser pointer and spirit level. A Rohde and Schwarz ZVH8 VNA was connected to the horn antennas for S21 measurements, with calibration performed by replacing the RIS tile with an aluminium plate of similar dimensions. The measured specular reflection magnitudes and phases for the 8 PIN diode configurations are plotted as the solid curves in Fig. \ref{phase_response}. It can be seen that the phase response follows the trend of the simulation relatively closely, with some divergence towards the higher end of the band, while the magnitude response is within 1 dB for most states.
\color{black}

In order to quantify the phase error versus frequency, the equivalent bit number has been employed in previous works with programmable metasurface reflectors discrete phase steps. Pereira et al. proposed a quality indicator for the phase response of discretely tuneable reflectarrays as the equivalent bit number \cite{Pereira2012}. The equivalent bit number is defined as:

\begin{equation}\label{NBIT}
    N_{bit} = \log_2 \left( \frac{360}{\sqrt{12}\sigma} \right)
\end{equation}

Where $\sigma$ is the phase standard deviation, in degrees, calculated from the set of phase differences, $\Delta\phi_m$, between the M discrete states:

\begin{equation}\label{PSD}
    \sigma = \sqrt{ \frac{\sum_{m = 1}^{M} (\Delta\phi_m)^3}{12\times360} }
\end{equation}

\begin{figure}
\centerline{\includegraphics[scale=0.6]{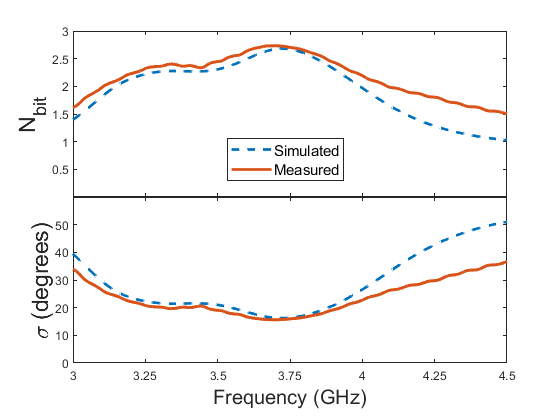}}
\caption{ Equivalent bit number (top) and phase standard deviation (bottom) for the 3-bit reflecting metasurface.}
\label{nbit_response}
\end{figure}

In their work, Pereira et al. introduced a 2-bit programmable reflectarray, and the useful bandwidth was considered where the phase resolution remains above 1.7 bits. This corresponds to a phase standard deviation lower than 32.5 degrees. There does not currently appear to be any convention for deciding the useful bandwidth beyond a 2-bit resolution. If we consider a phase standard deviation of 16.25 degrees (i.e., half that of a 2-bit design due to having double the number of states), this corresponds to an equivalent bit number of approximately 2.5 bits. Referring to Fig. \ref{nbit_response}, the design presented here can be seen to exhibit a phase standard deviation below 16.25 degrees in the frequency range of 3.62 to 3.78 GHz.

\color{black}

\section{Optimisation Algorithm}

The simple RIS optimisation algorithm presented here has been designed to find an RIS configuration which maximises the receiver power at a fixed location. When a VLoS link between the Tx and Rx via the RIS is established, the dominant path can be considered a combination of two LoS paths. That is, the LoS path between the transmitter and the RIS, and the LoS path between the RIS and the receiver. We adopt here the physics-compliant path loss model recently verified by Tang et al. \cite{Tang2019a} to approximate the received power at the Rx for a given RIS configuration and Tx position. The model assumes any LoS link between the Tx and Rx is negligible. The received power, $P_r$, at the position of Rx via the $N \times M$ set of RIS elements, with unit cell dimensions $d_x \times d_y$, can be approximated by:

\begin{multline}
\label{rxpower}
P_r = P_t\frac{G_tG_rd_xd_y\lambda^2}{64\pi^3} \\
\times\bigg|\sum_{m=1-\frac{M}{2}}^{M/2}\sum_{n=1-\frac{N}{2}}^{N/2}\frac{\sqrt{F_{n,m}^{combine}}\ \Gamma_{n,m}}{r_{n,m}^tr_{n,m}^r}e^{(-j\frac{2\pi}{\lambda}(r_{n,m}^t+r_{n,m}^r))}\bigg|^2
\end{multline}

\noindent with $P_t$ the transmit power, $G_t$ and $G_r$ the respective Tx and Rx antenna gains, $r_{n,m}^t$ and $r_{n,m}^r$ the distance between unit cell $\left(n,m\right)$ and the transmitter and receiver, respectively \cite{Tang2019a}. The term $F_{n,m}^{combine}$ takes into account the angle-dependent nature of the radiation patterns of the transmitter, unit cell reception, unit cell reradiation, and receiver, respectively.

%\begin{multline}
%\label{patterns}
%F_{n,m}^{combine} = %F^{tx}\left(\theta_{n,m}^{tx},\phi_{n,m}^{tx}\right)F\left(\theta_{n,m}^t,\phi_{n,m}^t\right) \\ \times F\left(\theta_{n,m}^r,\phi_{n,m}^r\right)F^{rx}\left(\theta_{n,m}^r,\phi_{n,m}^r\right)
%\end{multline}

The derivation of this channel model and similar recent advances in physics-compliant RIS channel models can be found in \cite{Tang2019a} and \cite{Esposti2021}, respectively. To maximise received power at the user equipment, the task is to find, within constraints of the operating environment (e.g., a wall-mounted RIS so as not to present an obstruction), the RIS dimensions, position, and set of unit cell bias states to maximise (\ref{rxpower}). For the case of the multi-bit RIS presented here, each set of column-connected unit cells is set to one of the 8 biasing configurations, $\rho_{k}e^{j\phi_{k}}$ where $k$ is an integer $k \in [1,8]$, $\rho_k$  and $\phi_k$ are the respective reflection magnitude and phase shift of configuration $k$. 

The algorithm adopted for these field trials is based on the adaptive optics-inspired approach employed by Gros et al. \cite{Gros2021}. For each controllable column of 12 unit cells, the set of 8 unit cell biasing configurations is iterated through and received power measurements are taken for each. The configuration which results in the highest received power is then selected for that column and the process continues for the remaining columns. This process is then repeated until received power improvement is negligible.

%% Start here today

\begin{algorithm}
\caption{Maximise Rx signal power}
\begin{algorithmic}[1]
\renewcommand{\algorithmicrequire}{\textbf{Input:} Average received signal power, $P_r$,}
\renewcommand{\algorithmicensure}{\textbf{Output:} RIS configuration, $\Gamma$,} %
\REQUIRE in
\ENSURE  out
%\textit{Initialisation} Initialise configuration matrix $ \boldmath{\Gamma} \in \mathBB{C}^{N \times M}$
\FOR {n = 1 to N}
    \FOR {m = 1 to M}
        \FOR {k = 1 to K}
        \STATE $ \boldmath{\Gamma_{n,m}} \leftarrow \rho_{k}e^{j\phi_{k}} $; 
        \STATE $ s_{k} = P_{r} $;
        \ENDFOR
        \STATE $q \leftarrow$ index of maximum of $s$;
        \STATE $ \boldmath{\Gamma_{n,m}} \leftarrow \rho_{q}e^{j\phi_{q}}$;
    \ENDFOR
\ENDFOR
\RETURN configuration matrix $\boldmath\Gamma$ 
\end{algorithmic} 
\end{algorithm}

\section{Indoor Field Trials}

For an existing network equipment deployment, such as a 5G New Radio (NR) small cell, it may be desirable to extend coverage to adjacent rooms, corridors, or floors without significant investment in additional infrastructure, such as a backhaul link and multiple RF chains \cite{Bjornson2020a}. The strategic placement of a RIS on interior walls could be a cost-effective solution to circumventing indoor blockages, in theory, but there are currently relatively few measurement campaigns in the literature to confirm this. Three indoor coverage enhancement scenarios have been considered in this work and are representative of situations where an additional small cell might usually be employed \cite{Hoppari2021}. Namely, an open lobby area adjacent to a common room, a junction between two long corridors, and a multi-floor scenario.

\subsection{Experimental Setup}

\begin{figure}
\centerline{\includegraphics[scale=0.2]{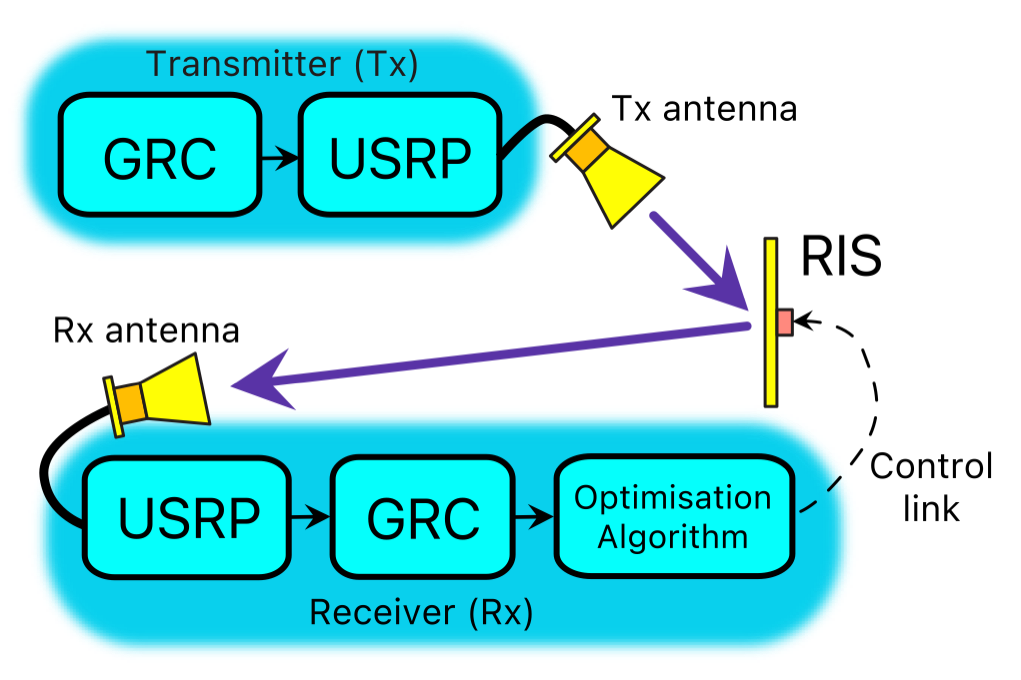}}
\caption{Diagram representing USRP-based communication link with RIS feedback loop. The RIS and receiver form a feedback loop via a WiFi control link to iteratively increase the received signal strength. }
\label{feedbackloop}
\end{figure}

The 5G NR standard employs OFDM on uplink and downlink for its high spectral efficiency and resilience to fading. In each experiment performed here, similar to previous field trial works \cite{Pei2021}\cite{Trichopoulos2021}, we utilised an OFDM signal with 20 MHz bandwidth and 312.5 kHz subcarrier spacing. The transmitted data was a randomly generated bit stream, with the signal processing performed by GNURadio Companion (GRC) software on laptop PCs. A block diagram of the RIS-aided communication link can be seen in Fig. \ref{feedbackloop}.

National Instruments Universal Software Radio Peripheral (USRP) X300 devices were used at the transmitter and receiver side. The USRP transmit power was fixed at 0 dBm and the receive gain set to 15 dB for all scenarios. The antennas employed were a monopole antenna with a gain of 2 dBi, as well as a pair of Aaronia PowerLOG 70180 standard gain horn antennas with a gain of approximately 10 dBi in the operating region of interest. These were arranged in a horizontal polarisation configuration in order to facilitate interaction with the horizontally polarised RIS. In each scenario, the RIS position and orientation were selected so as to be parallel to an adjacent wall, emulating wall-mounted indoor deployment. 

%%% Power settings of USRPs. 15 dB Rx gain. Tx power 5 dBm ? 

The measurements were performed as follows. All PIN diodes of the RIS were initially set to the off (000) state. The receiver was placed with its antenna directed towards the RIS with the aid of a laser pointer. The transmitter continuously transmits the aforementioned OFDM-modulated signal towards the RIS, and the receiver continuously samples the average squared magnitude of the received signal via GRC. These power measurements are polled by the RIS optimisation algorithm and the connected columns were iteratively optimised until 5 iterations of the algorithm had passed. The received power was then averaged over 1 second intervals for 2 minutes. In order to ascertain a reference case, the measurements were repeated with the RIS replaced by an aluminium plate of similar dimensions.

%% USRPs and antennas
%% Flowgraph structure

%% Algorithm 

\subsection{Scenario I - Lobby} 

%% Description of setup and Why scenario might be interesting
%% What happened
%% Why it might have happened

\begin{figure}
\centerline{\includegraphics[scale=0.13]{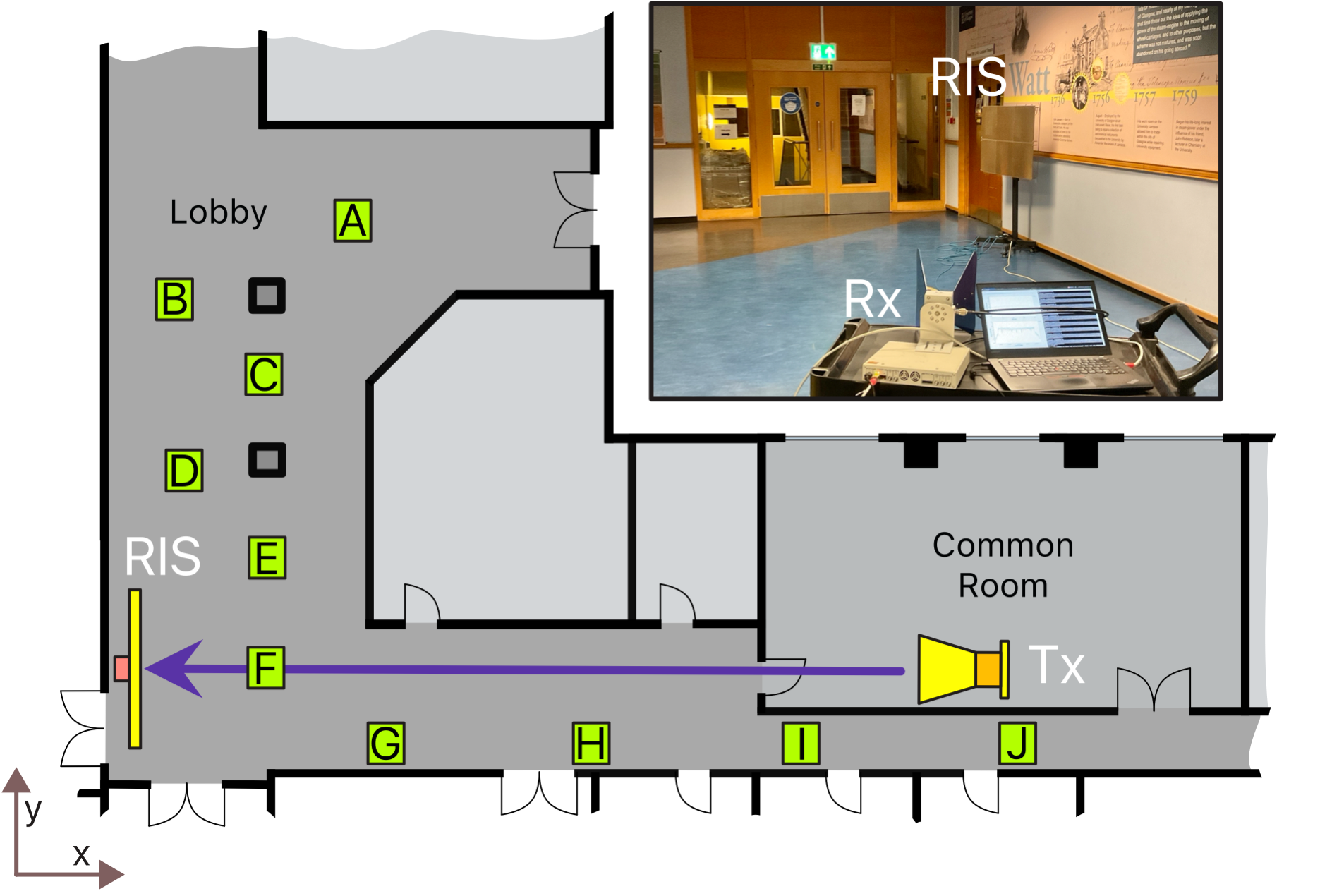}}
\caption{Experiment setup for scenario I. RIS in a lobby, with the transmitter (Tx) placed in an adjacent room, pointing via a doorway towards the RIS at a distance of 20 m. Locations A to J indicate receiver positions. Regarding the coordinates listed in table \ref{table_scene1}, the RIS is considered point (0, 0) and the transmitter at point (20, 0). The inset photo shows the receiver setup at position D.}
\label{scene_lobby}
\end{figure}

Scenario I is depicted in Fig. \ref{scene_lobby}. The transmitter USRP connected to a horn antenna is placed in the \textit{common room} at a height of 1 m and is directed towards the RIS placed in the \textit{lobby}, forming a LoS link. The RIS is positioned at a height of 1 m to its centre and, utilising algorithm 1, was optimised to maximise the received power at 10 different locations within the lobby, denoted A to J, the coordinates of which can be found in table \ref{table_scene1}. The origin (0, 0) is considered at the centre of the RIS and the Tx is positioned at (20, 0) (i.e., 20 meters broadside to the RIS). This setup is a typical indoor coverage extension scenario, with some locations benefiting from a strong LoS link from the transmitter (i.e., positions G and F). On the other hand, the path towards positions A, B, and C results in attenuation and scattering via propagation through the adjacent rooms, as well as potentially not being served by the main lobe of the transmit horn antenna. 

\begin{table}[!t]
\renewcommand{\arraystretch}{1.3}
\caption{Received power improvement for scenario I \\ (versus aluminium plate)}
\label{table_scene1}
\centering
\begin{tabular}{|l|l|l|l|}
\hline
\hline
\multicolumn{2}{|c|}{Location} &
\multicolumn{2}{c|}{Rx Power Improvement (dB)} \\
\hline
Identifier & Coordinates (meters) & Monopole & Horn \\
\hline
A & (5, 11) & 4.27 & 17.31 \\
\hline
B & (1, 8.5) & 8.93  & 17.11  \\
\hline
C & (3, 6.4) & 9.95 & 17.43  \\
\hline
D & (1, 4.6) & 7.01 & 15.47  \\
\hline
E & (3, 2.5) & 9.53  & 17.48  \\
\hline
F & (3, 0) & 16.10 & 9.12  \\
\hline
G & (6, -2) & 13.39 & 14.51  \\
\hline
H & (11, -2) & 3.97 & 7.47 \\
\hline
I & (17, -2) & 0.21 & 1.51  \\
\hline
J & (23, -2) & 2.54 & 5.04 \\
\hline
\hline
\end{tabular}\end{table}

Two different antennas were deployed at the receiver in order to compare the received power improvement performance, namely, a  monopole antenna and a standard gain horn antenna. The monopole was found to exhibit acceptable (i.e., -10dB) return loss around 3.9 GHz which is within the operating region of the fabricated RIS, and therefore the experiment was performed at this frequency. The receiver PC, USRP, and antennas were placed on a 1m high trolley, as depicted in the photo in Fig. \ref{scene_lobby}.  The receiver was placed in one of the 10 locations throughout the lobby, with the receive antenna directed towards the RIS with the aid of a laser pointer.

%The measurements were performed as follows. All PIN diodes of the RIS were initially set to the off (000) state. The receiver was placed in one of the 10 locations throughout the lobby in Fig. \ref {scene_lobby}, with the receive horn antenna directed towards the RIS with the aid of a laser pointer. The transmitter continuously transmits the aforementioned OFDM-modulated signal towards the RIS, and the receiver continuously samples the average squared magnitude of the received signal. These power measurements are polled by the RIS optimisation algorithm ## and the connected columns are iteratively optimised until convergence. The received power was then averaged over 1 second intervals for 2 minutes in order to mitigate the effects of any interference. In order to ascertain a reference case, the measurements were repeated with the RIS replaced by an aluminium plate of similar dimensions. 

The resulting power improvement over the metal plate for the monopole and horn antenna cases is shown in table \ref{table_scene1}. The horn antenna placed at positions A to E appears to benefit most from the introduction of the RIS due to the blockage caused by the adjacent rooms. The notable difference in improvement between the monopole and horn cases is due to the differing directivity of these antennas. Prior to optimisation of the RIS, the monopole receives power via NLoS paths from a more diverse set of directions compared to the horn antenna, resulting in a larger apparent improvement. Position I does not significantly benefit from the RIS in the monopole case and only marginal improvement of 1.51 dB for the horn case. This is likely due to the proximity of the transmitter to the adjacent wall, resulting in a much stronger non-LoS signal path over the short distance.

\subsection{Scenario II - Corridor Junction}

%% Why scenario might be interesting
%% Description of setup
%% What happened
%% Why it might have happened

\begin{figure}
\centerline{\includegraphics[scale=0.11]{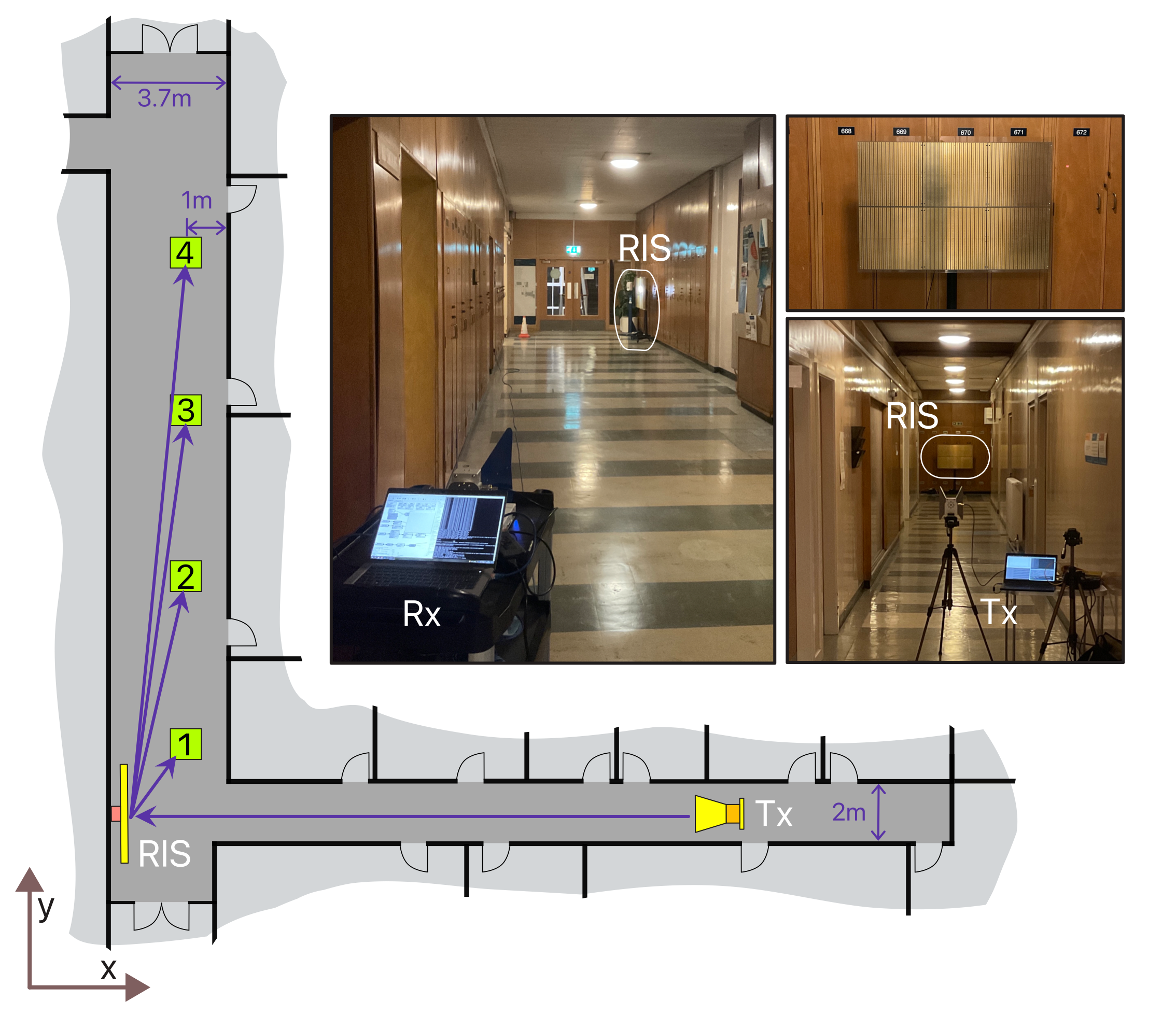}}
\caption{Experiment setup for scenario II. Transmitter is placed in a connecting corridor, pointing directly towards the RIS at a distance of 20 m. Locations 1 to 4 indicate receiver positions. Regarding the coordinates listed in table \ref{table_scene2}, the RIS is considered point (0, 0), with \textit{x} broadside to the surface. }
\label{scene_corridor}
\end{figure}

In this experiment, a similar system setup to scenario 1 was adopted via a corridor junction, as depicted in Fig. \ref{scene_corridor}. The RIS was placed at the junction between a wide and narrow corridor, with the surface normal to the narrow corridor, and can be considered point (0, 0). The transmitter, directed towards the RIS, was placed in the narrow corridor at a distance of 20 m (20, 0) from the RIS. Four receiver locations were selected, denoted positions 1 to 4, placed at 5 m intervals along the y axis, starting from (2, 2). The coordinates of the receiver positions alongside the resulting received power improvement at 3.9 GHz can be found in table \ref{table_scene2}. It can be seen that there is notable received power improvement in both the monopole and horn antenna cases, with up to 30.6 dB improvement at position 4. %% Is there an explanation for position 3? and why do we get 30.6 dB improvement at 4 for horn but only 11.8 dB for monopole? Why do the trends differ?

For positions 1 and 3, the received power improvement appears to exhibit similar behaviour for both monopole and horn cases, whereas there is a significantly larger improvement for the horn antenna case at positions 2 and 4. Receivers served via NLoS propagation typically experience a wide power angle spectrum compared to LoS scenarios \cite{Kafle2008}. We can therefore expect the monopole antenna to more consistently capture power in a rich scattering environment since its peak antenna gain direction is less affected by its orientation. This is reflected in the results shown in Table \ref{table_scene2}, where the horn antennas achieve a significantly higher relative improvement compared to the monopole case. In their indoor channel sounding work, Wallace et al.,  \cite{Wallace2017} observed a more favorable channel gain performance in indoor NLoS scenarios whilst utilising a monopole array as opposed to a more directional patch antenna array. They posited it is likely due to the reduced multipath richness in the patch array case. Similarly, we believe that the smaller relative improvement in the monopole case for locations 2 and 4 is due to the fact that the monopole was receiving significantly more power in the reference case at those locations compared to the more directional horn antennas. Additionally, during reference measurements the horn antennas were directed at the aluminium reference plate whose scattering pattern exerts a greater influence on the received power than the monopole case due to the narrow field of view of the horn. We believe the combination of the rich scattering environment and the variation in the nulls and sidelobes from the reference plate with respect to the receiver position results in the apparent improvement of the horn antenna exhibiting notably more variation than the monopole case.

\color{black}

\begin{table}[!t]
\renewcommand{\arraystretch}{1.3}
\caption{Received power improvement for scenario II \\ (versus aluminium plate) for $f = 3.9\ GHz$}
\label{table_scene2}
\centering
\begin{tabular}{|l|l|l|l|}
\hline
\hline
\multicolumn{2}{|c|}{Location} &
\multicolumn{2}{c|}{Rx Power Improvement (dB)} \\
\hline
Identifier & Coordinates (meters) & Monopole & Horn \\
\hline
1 & (2, 3) & 20.41 & 22.23 \\
\hline
2 & (2, 8) & 15.70 & 27.68 \\
\hline
3 & (2, 13) & 15.9 & 16.84 \\
\hline
4 & (2, 18) & 11.8 & 30.60 \\
\hline
\hline
\end{tabular}\end{table}

\begin{figure}
\centerline{\includegraphics[scale=0.36]{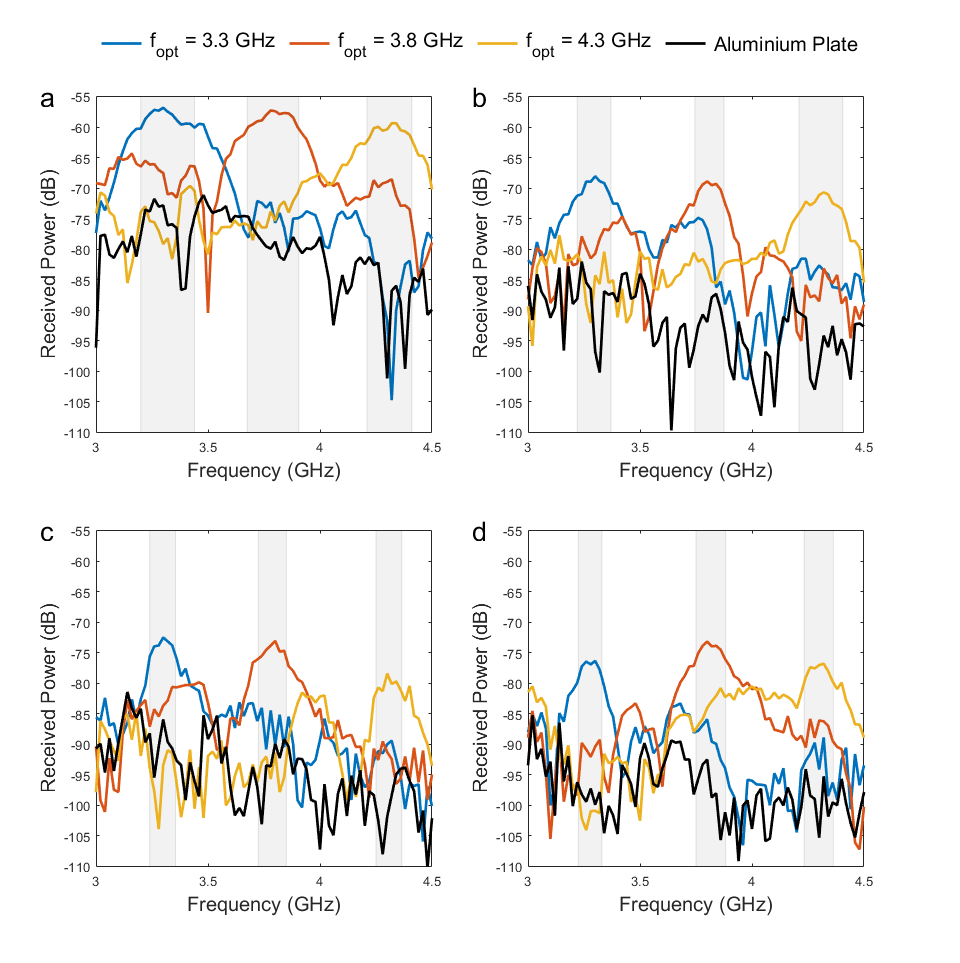}}
\caption{Received power versus frequency for scenario II. (a) to (d) show received power at positions 1 to 4, respectively. Optimisation of the RIS was performed at 3.3 GHz, 3.8 GHz, and 4.3 GHz, with a clear improvement over the metal plate of the same dimension. In plot (a) it can be seen that a received power improvement as much as 40 dB is achieved in the 4.3 GHz optimised configuration.}
\label{fsweep_scene2}
\end{figure}

\begin{figure}
\centerline{\includegraphics[scale=0.38]{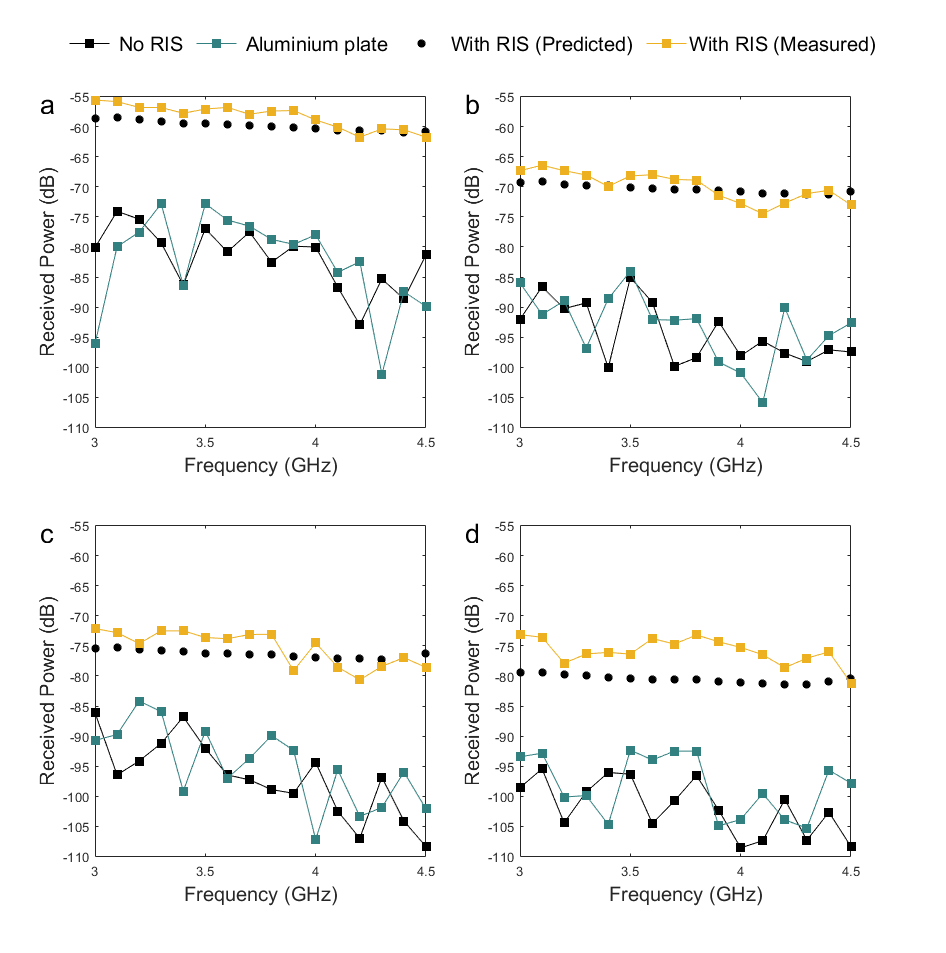}}
\caption{Envelope of the full set of curves from Fig. \ref{fsweep_scene2}. Maximum received power in scenario II at positions 1 to 4 for steps of 100 MHz for the cases of no RIS, the aluminium reference plate, the path loss model of (\ref{rxpower}) for an optimised RIS, and measured optimised RIS. Received power improvement can clearly be achieved at intervals across the 1.5 GHz range.}
\label{fopt_scene2}
\end{figure}

In order to investigate the useful bandwidth of the RIS in this scenario, the experiment was repeated with the horn antenna at the receiver between 3 and 4.5 GHz at 100 MHz intervals. At each location, the RIS was optimised for a single frequency point, $f_{opt}$, followed by performing a sweep from 3 to 4.5 GHz at 20 MHz intervals and measuring the average received power for each interval. This was then repeated to obtain 16 frequency sweeps per location. The resulting received power versus frequency for $f_{opt} = 3.3$, $3.8$, and $4.3$ GHz, as well as for the aluminium plate, can be seen in Fig. \ref{fsweep_scene2}. The 3 dB bandwidths for each case are highlighted by grey bars. Referring to Fig. \ref{fsweep_scene2}, the aluminium plate case can be seen to be highly frequency-selective, as can be expected for a NLoS scenario. This selectivity can be eliminated over finite regions of the band when the RIS is optimised at the carrier frequency of interest. 

The resulting maximum power improvement for each $f_{opt}$ have been plotted in Fig. \ref{fopt_scene2}, where the aluminium plate case and the case of no RIS have been plotted for comparison, as well as the predicted Rx power from (\ref{rxpower}). It can be seen that the introduction of the RIS offers selective improvement across the entire 40\% bandwidth of the investigation, with each configuration offering between 105 and 240 MHz instantaneous 3dB-bandwidth. Equation (\ref{rxpower}), subject to algorithm 1, can be seen to offer an excellent prediction of the maximum achievable power in this scenario for the cases of positions 1 to 3. 

%% Monopole benefits from a larger diversity of paths between Rx and Tx.  

\subsection{Scenario III - Floor to Floor}

%% Should put magnitude here too
\begin{table}[!t]
\renewcommand{\arraystretch}{1.3}
\caption{ RIS Reflection Behaviour and  \\ Received Power Improvement for Scenario III 
\\ (versus aluminium plate)}
\label{table_multibit}
\centering
\begin{tabular}{|l|l|l|l|l|l|}
\hline
\hline
\multicolumn{3}{|c|}{State} &
\multicolumn{3}{c|}{Resolution} \\
\hline
Config & $|\Gamma|$ & $\angle\Gamma$  & 1-Bit & 2-Bit & 3-Bit \\
\hline
011 & -0.77 dB  & $-180^{\circ}$     &  \cellcolor{gray!90}  & \cellcolor{gray!70} & \cellcolor{gray!30} \\
\hline
111 & -0.50 dB  & $-174.1^{\circ}$   &                  &                  & \cellcolor{gray!30} \\
\hline
110 & -1.60 dB  & $-133.8^{\circ}$  &  &                  & \cellcolor{gray!30} \\
\hline
101 & -1.43 dB  & $-92.9^{\circ}$  &                  & \cellcolor{gray!70} & \cellcolor{gray!30} \\
\hline
010 & -2.91 dB  & $-59.2^{\circ}$ &                  &                  & \cellcolor{gray!30} \\
\hline
001 & -3.05 dB  & $1.4^{\circ}$ & \cellcolor{gray!90}  & \cellcolor{gray!70} & \cellcolor{gray!30} \\
\hline
100 & -0.11 dB  & $43.9^{\circ}$ &  &                  & \cellcolor{gray!30} \\
\hline
000 & -0.72 dB  & $122.2^{\circ}$ &                  & \cellcolor{gray!70} & \cellcolor{gray!30} \\
\hline
\multicolumn{3}{|c|}{Rx Power Improvement (dB)} & 18.48 & 20.45 & 21.13 \\
\hline
\hline
\end{tabular}\end{table}

\begin{figure*}[htbp]
\centerline{\includegraphics[scale=0.15]{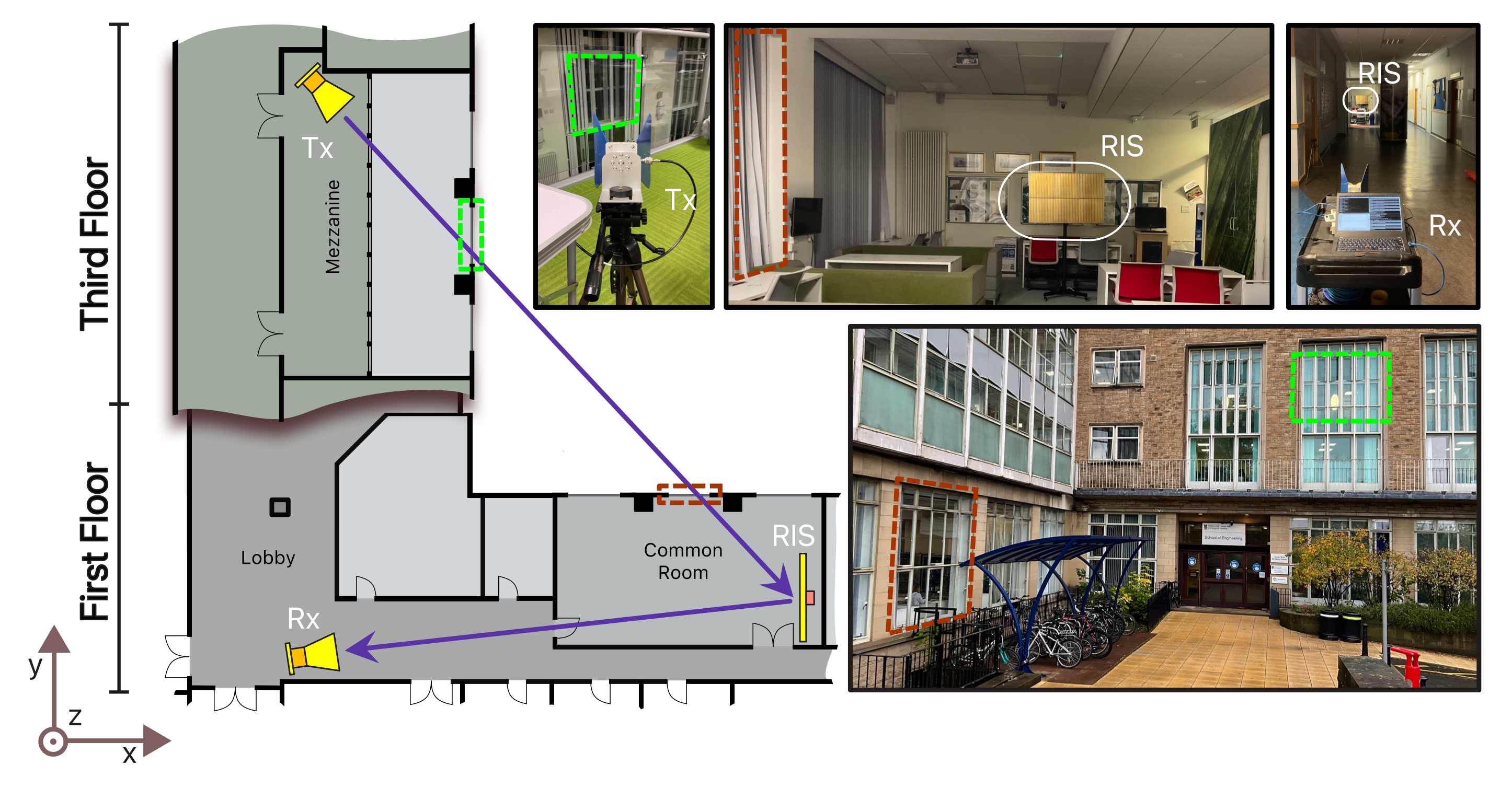}}
\caption{Experiment setup for scenario III. The RIS is placed in the common room of scenario I, serving a Rx placed in the lobby. A transmitter is located two floors above, on a mezzanine with a direct path to the RIS via the two windows highlighted bottom-right. The receiver (Rx) is considered point (0, 0, 0), with the transmitter (Tx) and RIS located at (20, 20, 10) and (20, 2, 0), respectively.}
\label{scene_multifloor}
\end{figure*}

\begin{figure}
\centerline{\includegraphics[scale=0.55]{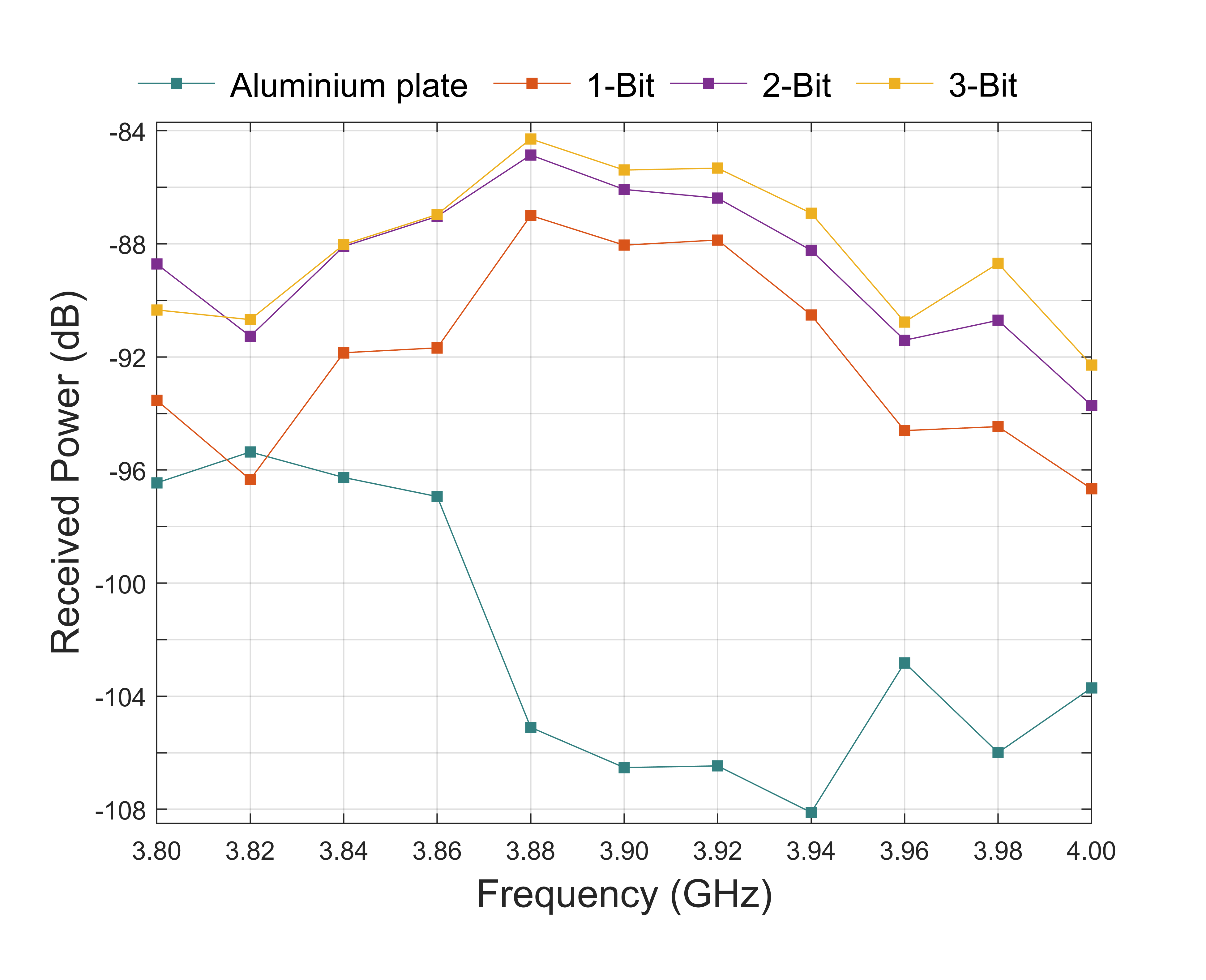}}
\caption{Received power versus frequency in scenario III after optimisation at 3.9 GHz. Received power for phase resolution settings of 1-bit, 2-bit, and 3-bit are plotted along with the reference aluminium plate for comparison.}
\label{fsweep_scene3}
\end{figure}

%% Why scenario might be interesting
%% Description of setup
%% What happened
%% Why it might have happened
Due to the column-connected architecture of the RIS design introduced here, wave transformation capability in elevation is limited when compared to designs where unit cells are individually addressable. This does not pose a significant problem in scenarios where both the electrical distance of antennas from the RIS is such that plane wave incidence can be approximated and when the antennas are located in the same horizontal plane as the RIS. For indoor communication scenarios, the latter case is quite likely to occur when user equipment is located on the same floor. In order to ascertain whether this RIS architecture can offer coverage improvement when the latter criteria is not met, a multi-floor experiment was devised and is depicted in Fig. \ref{scene_multifloor}. 

The receiver was placed in the first floor lobby from scenario 1, whilst the RIS was placed in the common room with a LoS link to the Rx horn antenna. The transmitter antenna was placed on a mezzanine two floors above, where a wireless link was formed to the RIS via two windows, highlighted by dashed rectangles. In order to avoid interference with a nearby small cell operating at 3.75 GHz, the carrier frequency was set to 3.9 GHz and a frequency sweep was performed over a 200 MHz bandwidth. 

So as to demonstrate the benefit of the high phase resolution of this RIS design, measurements were performed for 1-bit, 2-bit, and 3-bit cases. The digital states available to the optimisation algorithm in each case are listed in table \ref{table_multibit}. 
The resulting received power has been plotted in Fig. \ref{fsweep_scene3}, where it can be seen that a 21.13 dB improvement over the aluminium plate case was possible in this scenario with the 3-bit phase resolution. When limiting the phase resolution to 2-bit and 1-bit, the optimised received power reduces by 0.68 dB and 2.65 dB, respectively. According to \cite{Wu2008}, the average directivity reduction for a large reflectarray subject to plane wave excitation compared to a 3-bit design is approximately 0.67 dB and 3.66 dB for 2-bit and 1-bit designs, respectively. However, this approximation does not take into account phase-dependent magnitude and instead assumes an ideal unity reflection magnitude. The combination of phase-dependent magnitude and limited phase range of this design compared to an ideal 3-bit case may explain the 1 dB difference between the measured directivity reduction in the 1-bit case to its theoretical value.

\section{Discussion}

%This section discusses the implications of the obtained measurements and provides a comparison with previously published results.

%%%%%%%%%%%%% PASTE TABLE CONTENT
\begin{table*}[]
\caption{ Comparison of RIS Field Trial Works at Sub-6 GHz}
\label{table_compare}
\begin{tabular}{|l|l|l|l|l|l|l|l|}
\hline
\hline
\textbf{Ref.}                       & \textbf{\begin{tabular}[c]{@{}l@{}}Phase\\ resolution\end{tabular}} & \textbf{\begin{tabular}[c]{@{}l@{}}Tuning\\ mechanism\end{tabular}}          & \textbf{\begin{tabular}[c]{@{}l@{}}Dimensions\\ No. elements\\ (Cols x Rows)\end{tabular}} & \textbf{Frequency} & \textbf{\begin{tabular}[c]{@{}l@{}}Investigated\\ bandwidth\end{tabular}} & \textbf{Scenarios}                                                                                         & \textbf{\begin{tabular}[c]{@{}l@{}}Effects on\\ channel response\end{tabular}}                                                                                                                                \\ \hline
\cite{Pei2021}          & 1-bit                                                               & \begin{tabular}[c]{@{}l@{}}Varactor \\ diodes\end{tabular} & \begin{tabular}[c]{@{}l@{}}0.80 m $\times$ 0.31 m
\\ ($ 15.5\lambda \times 6\lambda $) \\ 1100 elements\\ (55 $\times$ 20)\end{tabular}        & 5.8 GHz            & 500 MHz                                                                   & \begin{tabular}[c]{@{}l@{}}Indoor NLoS,\\ Outdoor rooftops \\ at 50 and 500 m.\end{tabular}                & \begin{tabular}[c]{@{}l@{}} \\ 27 dB and 14 dB received power\\ improvement for 50 m and\\ 500 m outdoor links, respectively.\\ 26 dB channel gain enhancement \\ indoors via a thick concrete wall. \\ \\ \end{tabular} \\ \hline
\cite{Trichopoulos2021} & 1-bit                                                               & PIN diodes                                                                   & \begin{tabular}[c]{@{}l@{}} \\ 0.41 m $\times$ 0.26 m \\ ($ 7.9\lambda \times 5\lambda $) \\ 160 elements\\ (16 x 10) \\ \\  \end{tabular}         & 5.8 GHz            & 150 MHz                                                                   & \begin{tabular}[c]{@{}l@{}}Outdoor NLoS via \\ large occlusion.\end{tabular}                      & \begin{tabular}[c]{@{}l@{}} \\ 6 dB average SNR improvement\\ throughout coverage blind spot. \\ \\ \end{tabular}                                                                                                        \\ \hline
\cite{246282}           & 1-bit                                                               & RF switches                                                                  & \begin{tabular}[c]{@{}l@{}}2.45 m $\times$ 2.45 m  \\ ($ 19.6\lambda \times 19.6\lambda $) \\ 3200 elements\\ (64 x 50)\end{tabular}        & 2.4 GHz            & 38 MHz                                                                   & \begin{tabular}[c]{@{}l@{}}Indoor office environment\\ with mixed LoS/NLoS\end{tabular}                    & \begin{tabular}[c]{@{}l@{}} \\ Median 9.8 dB signal level\\ improvement and doubling of\\ channel capacity. \\ \\  \end{tabular}               

\\ \hline

\cite{Dan2022}          & 1-bit                                                               & PIN diodes                                                                   & \begin{tabular}[c]{@{}l@{}}- \\ 512 elements\\ (16 x 32)\end{tabular}                       & 2.64 GHz           & 160 MHz                                                                   & \begin{tabular}[c]{@{}l@{}}Indoor NLoS at a \\ corridor junction.\end{tabular}                               & \begin{tabular}[c]{@{}l@{}} \\ 10 dB signal level improvement\\ compared to reference of RIS\\ with random configuration. \\ Demonstrated 10 Mbps\\ throughput increase. \\ \\ \end{tabular}                             \\ \hline

\cite{Araghi2022}       & Continuous                                                          & \begin{tabular}[c]{@{}l@{}}Varactor \\ diodes \end{tabular}                                                              & \begin{tabular}[c]{@{}l@{}}\\ 1.14 m $ \times $ 1.16 m  \\ ($ 13.3\lambda \times 13.5\lambda $) \\ 2430 elements\\ (30 x 81) \\ \\\end{tabular}        & 3.5 GHz            & -                                                                         & \begin{tabular}[c]{@{}l@{}}Indoor NLoS. Room to \\ room wireless link.\end{tabular}                         & \begin{tabular}[c]{@{}l@{}} \\ 15 dB received signal strength\\ enhancement versus RIS \\ unconfigured. \\ \\ \end{tabular}                                                                                                                          \\ \hline

\begin{tabular}[c]{@{}l@{}} This \\ work       \end{tabular}                          & 3-bit                                                               & \begin{tabular}[c]{@{}l@{}}PIN diodes\\ \end{tabular}      & \begin{tabular}[c]{@{}l@{}}1.02 m $ \times $ 0.72 m \\ ($ 12.8\lambda \times 9\lambda $)\\ 2304 elements\\ (48 x 48)\end{tabular}        & 3.75 GHz           & 1.5 GHz                                                                   & \begin{tabular}[c]{@{}l@{}} Indoor mixed LoS/NLoS.\\ Corridor junctions and\\ multiple floors.\end{tabular} & \begin{tabular}[c]{@{}l@{}} \\ Average 16 dB received signal \\ strength enhancement over\\ entire 1.5 GHz bandwidth in \\ corridor junction. \\ Up to 40 dB signal level \\ improvement in deep fading. \\ Multi-floor signal strength \\ enhancement above 20 dB \\ \\ \end{tabular}                     \\ \hline
\hline
\end{tabular}
\end{table*}

In order to contrast the findings of this work with previous observations, we have compiled the information in Table \ref{table_compare}. It is difficult to draw direct comparisons with similar realistic RIS field trial works due to the large variations in operating environments, RIS dimensions, and control degrees of freedom. As can be seen on the right-hand column of Table \ref{table_compare}, a common trend is the promising signal strength enhancement in each case. These signal strength metrics are usually given with a reference case, as we have employed in this work, and the resulting difference between the reference and optimised RIS measurements can vary profusely. This is particularly the case when the receiver is in close proximity to the RIS as in the indoor scenario in \cite{Pei2021}. 

Performance of RISs beyond a narrow bandwidth is not widely documented and we aimed here to address this. In comparison to the 1-bit cases in previous works, it was possible in this work to achieve notable channel improvement well beyond the proximity of the design frequency due to the dispersive nature of the many RIS states. For example, on observing the measured global reflection response depicted in Fig. \ref{phase_response}, if only two states were available, it can be seen that no two states can be selected that would provide a 1-bit response across the 1.5 GHz bandwidth. For example, if we were limited to the 001 and 011 states shown in Fig. \ref{phase_response}, we could achieve a 1-bit response at 3.75 GHz. However, the global reflection behaviour would degenerate into less than 50 degrees relative phase difference at 3.25 GHz and 4.25 GHz, resulting in the RIS having a much more limited effect on the channel. However, the introduction of additional states in this multi-bit RIS enables at the very least a 1-bit response for the entire band and is capable of a 2-bit response between 3.2 and 4.1 GHz. It would be possible to extend this favorable reflection behaviour across a wider bandwidth by optimising the metasurface geometry for a flat equivalent bit number as opposed to the maximum value opted for at 3.75 GHz. Such modifications may not have notably affected scenario 3, where it can be seen that only a 0.68 dB improvement is offered for the 3-bit case over the 2-bit.

%% Scenario compare

In each scenario here, we have positioned the RIS where it might reasonably be installed. In reality, we could see the deployment of these devices with little noticeable change to the observable environment, such as on the facades on buildings and embedded within walls and ceilings. This is in contrast to the corridor scenario in \cite{Dan2022} where the RIS presents a clear obstruction to personnel and the indoor through-wall scenario in \cite{Pei2021} where the RIS was placed in the middle of a room. To ensure a realistic look at RIS capabilities, in this investigation we positioned the RIS to always be parallel to the interior walls of the building. Amongst the deep fading measured at location 1 in the corridor junction scenario, we achieved as much as 40 dB received signal strength improvement. This is in league with the 26 dB channel gain enhancement observed by Pei et al. \cite{Pei2021} in their indoor measurements with the RIS and transmitter completely isolated by a 30cm-thick concrete wall, and a receiver horn antenna at a similar number of wavelengths from location 1. In comparison to the work by Araghi et al. \cite{Araghi2022} employing their continuously-tunable RIS of a marginally larger dimension relative to the wavelength, we witnessed a similar received signal power improvement in our indoor corridor scenario with a RIS of smaller size. However, continuously tunable metasurfaces are much more complex, especially for local control, requiring a digital to analog converter channel for each element or grouping of elements. Our investigation has shown that similar channel enhancement could be achieved whilst utilising digital circuitry in combination with a monoplanar (i.e., vialess) metasurface. This reduces the implementation and fabrication complexity whilst maintaining a desirably high phase resolution.

In the multi-floor experiment, we have compared phase resolution versus channel improvement via our multi-bit RIS. There is notable improvement as the resolution is increased beyond 1-bit, but this is a comparison of the same device with varying degrees of freedom. Therefore, these results should be carefully interpreted. A 1-bit RIS only requires a phase range of $ 180^\circ $, as opposed to 2- and 3-bit RISs requiring phase ranges of $ 270^\circ $ and $ 315^\circ $, respectively. To achieve a higher resolution, a deeper resonance is required in order to provide a greater phase difference between the upper and lower phase limits of the local reflection responses. This inevitably results in higher losses due to the increased local electric field strength associated with a strong resonant response \cite{Qu2015}. For example, the 1-bit monoplanar unit cell design recently introduced by Trichopoulos et al. \cite{Trichopoulos2021} exhibited a 1 dB average reflection loss at the centre frequency as opposed to the 1.8 dB average reflection loss in the two states utilised for scenario 3 in our work. We would therefore expect a slightly higher achievable power level with a RIS specifically designed to operate with a 1-bit phase resolution for a similar RIS size and degree of local control.

%% What else could this mean for industry?
%% Consider what 3-bit could mean compared to 1-bit

%%%%%%%%%%%%% END PASTE TABLE CONTENT

%% Implications for real-world

As we have demonstrated in the diverse indoor settings explored here, RISs have the capability to mitigate the coverage holes plaguing existing mobile networks without introducing more radiation into the congested sub-6 GHz wireless spectrum. All power gains observed in this work arose purely from the optimal biasing of a number of PIN diodes with a DC voltage. Of particular interest to mobile operators may be the multi-floor experiment, which exhibits similarities with an outdoor-to-indoor coverage enhancement scenario common to urban environments. Combining scenarios 1 and 3, with multiple RISs, an operator could reach users in the lobby area, deeper inside the building, and maintain higher quality of service due to the 100-fold received signal strength improvement observed in scenario 3. Alternatively, transmit power at the base station could be reduced proportionally, resulting in operating cost reductions and reduced inter-cell interference. Any power consumption reduction prediction, however, must take into account the additional overhead incurred throughout the optimisation process, such as during computation and maintaining a control link, as well as any power consumed by the RIS. 

The minimum observed instantaneous bandwidth in the corridor junction and multi-floor scenarios was above 100 MHz, encompassing all channel bandwidth settings of widely deployed 5G systems at mid-band \cite{bertenyi20185g}. From this investigation it can be seen that a single RIS architecture engineered to exhibit a wide operational bandwidth could potentially serve the entire 5G N77 3.3 - 4.2 GHz band.

\section{Conclusion}

The paper presents indoor field trials using a 3-bit azimuthal-steering RIS. The wealth of literature on RIS channel models and optimisation algorithms should be complimented by rigorous field trials in realistic communication scenarios. The high phase resolution reconfigurable metasurface introduced and employed in this work shows significant NLoS channel improvement over and beyond the 3.4 to 3.8 GHz 5G commercial services band. \color{black} Indoor received power improvement of up to 40 dB with an instantaneous bandwidth of at least 100 MHz has been realised, easily meeting the demands of network providers for time-division duplexing operation at sub-6 GHz. The benefits of increased phase resolution have been demonstrated in a multiple-floor channel improvement experiment, where nearly double the power is received for the same RIS surface area by inclusion of additional reflection phase states. The RIS design explored here demonstrates that we do not necessarily require individually addressable unit cells to greatly benefit from this technology.

Future works might explore more complex groupings of unit cells than  a column-connected approach in order to further reduce the control network complexity, potentially rendering large RISs feasible in practice.

%RISs that are tailored for a particular operating environment could result in lower RIS complexity by removing redundant wave transformation capability. The RIS introduced here may be unsuitable in situations where the transmitter and receiver are bbbbbbbbb located in the same horizontal plane, but might be successfully deployed outdoors in scenarios where the base station and UE are located at equal and opposite angles relative to the elevation plane, thereby users could be served by a combination of specular reflection in the vertical direction and anomalous reflection azimuthally. 

% or


%\appendices

%\section{}

\section*{Acknowledgment}
James Rains' PhD is funded by EPSRC ICASE studentship (EP/V519686/1) with British Telecom.
Tie Jun Cui acknowledges support from the National Key Research and Development Program of China (2017YFA0700201, 2017YFA0700202, and 2017YFA0700203). \\

\ifCLASSOPTIONcaptionsoff
  \newpage
\fi



% http://www.michaelshell.org/tex/ bibtex/
%\bibliographystyle{IEEEtran}

\bibliography{bib1}
\bibliographystyle{ieeetr}


%\begin{IEEEbiography}{James Rains}
%\end{IEEEbiography}


%\newpage

%\begin{IEEEbiographynophoto}{James Rains}


%\vfill


\end{document}